\documentclass[11pt,twoside]{article}

\usepackage{asp2006}
\usepackage{epsf}
\usepackage{psfig}
\usepackage{lscape}

\begin{document}
\title{Orbital parameters of binary radio pulsars in globular clusters and stellar interactions}   
\author{Manjari Bagchi and Alak Ray}
\affil{Tata Institute of Fundamental Research, Mumbai, India 400005}

\begin{abstract} 
The observed distribution of globular cluster binary radio pulsars in the eccentricity versus orbital period plane can be explained as a result of binary-single star
interactions. Our numerical and analytical study hints that the highest eccentricity binaries in clusters are likely to be from exchange and/or merger of a single star with a binary component, while the intermediate eccentricity systems are probably results of fly-by interactions.

\end{abstract}

\section{Introduction}
\label{sec:intro}

Observable parameters of neutron star (NS) binaries in globular clusters (GC), such as spin period ($P_{spin}$), orbital period ($P_{orb}$), eccentricity ($e$), projected radial position in the cluster, companion mass  $etc.$ provide a
valuable test-bed to examine the theoretical scenarios of
formation and evolution. These parameters can even be used as a tracer of the past history of dynamical interactions of the binary NSs in individual GCs. As orbital parameters of LMXBs are poorly known, we concentrate entirely on binary radio pulsars whose orbital parameters can be determined accurately by pulsar timing. We study the distribution of GC pulsars in $e-P_{orb}$ plane as a result of  interactions with single stars. 

\section{Observed data}
\label{sec:data}

So far 140 pulsars in 26 GCs\footnote{http://www.naic.edu/$\sim$pfreire/GCpsr.html} have been discovered, 74 are binaries, 59 are isolated and 7 without published timing solutions. PSR J2140$-$2310B with poorly known orbital parameters is excluded from the present study. Majority of the rest 73 binary pulsars are millisecond pulsars whose $e$ Vs $P_{orb}$ distribution is shown in Fig \ref{fig:pulsars_all_group}. They can be categorized into three groups: I) 21 pulsars with large eccentricity ($1 > e \geq 0.01$); II) 20 pulsars with moderate eccentricity ($0.01 > e \geq 2 \times 10^{-6}$); III) 32 pulsars with small eccentricity ($e=0$ in the database, but $e~=~3 \times 10^{-7}$ in present work).  The smallest measurable value of $e$ follows the relation \cite{phi92}: $e_{min}= (\delta t) / (a sin~i /c) = 4 \pi^2 c \delta
t/ \left[sin~i \left(G(m_p+m_c)^{1/3}\right)
P_{orb}^{2/3}\right]$. We take $m_p~{\rm (pulsar~ mass)}$ $=~1.4~M_{\odot}$,
$m_c~{\rm (companion~ mass)}=~0.35~M_{\odot}$, $i~{\rm (inclination ~angle)}=60^{\circ}$, $\delta t$ ${\rm (timing~ accuracy)}$ $=~ 1 ~\mu{\rm
sec}$ to draw this ``limit of timing sensitivity" line (the lower left corner of Fig. \ref{fig:pulsars_all_group}). Observational selection effect may be influencing the distribution in the low $P_{orb}$ region (Camilo \& Rasio 2005).

\section{Stellar interactions in globular clusters}
\label{sec:interact}

\begin{figure}[h!]
\plotone{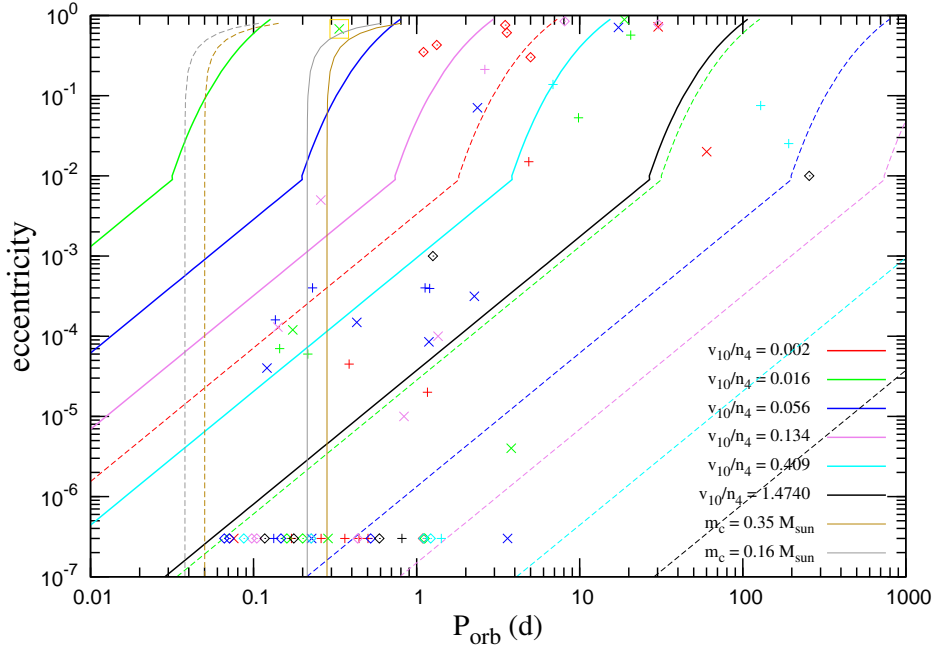}
\caption{GC binary pulsars in the $e-P_{orb}$
plane with contours of $t_{fly}=10^{10}$ yrs (solid lines) and
$t_{fly}=10^{8}$ yrs (dashed lines) for different values of
$v_{10}/n_{4}$ and contours of $t_{gr}=10^{10}$ yrs (solid lines)
and $t_{gr}=10^{8}$ yrs (dashed lines) for binaries with
$m_p~=~1.4~ M_{\odot}$ and $m_c~=~0.35~ M_{\odot}$ or $0.16~
M_{\odot}$. The color scheme for
$t_{fly}$ and $t_{gr}$ contours are shown on the right.
Individual pulsars are marked with same colors as
$v_{10}/n_{4}$ values of their host GCs. Pulsars with projected positions inside the
cluster core are marked with $+$, those outside the cluster core
with $\times$ and the pulsars with unknown positions with
$\diamond$. Six binaries in Terzan 5 with $e > 0.1$ are marked with their names. \label{fig:pulsars_all_group}}
\end{figure}

Due to high stellar densities, binary - single star interaction is significant in GCs resulting any one of among ``fly-by" , ``exchange" , ``merger" or  ``ionization". Fly-by, exchange and merger interaction can produce eccentric binary millisecond pulsars when the interaction timescale is less than the binary age wheras formation scenario of millisecond pulsars suggests that their eccentricity should be very small, $\sim 10^{-6}-10^{-3}$ (Phinney 1992). We take the maximum age of the binaries in a GC to be equal to GC ages ($\sim 10^{10}$ yrs). Rasio \& Heggie (1995) obtained the expressions for fly-by time-scales for an initially circular binary as: $$t_{fly}=4\times 10^{11}
n_4^{-1}v_{10}P_{orb}^{-2/3}e^{2/5}~ \rm {~for ~e \leq 0.01}$$
$$t_{fly}=2\times 10^{11}
n_4^{-1}v_{10}P_{orb}^{-2/3}\left[- \ln(e/4) \right]^{-2/3} ~\rm
{~for~e \geq 0.01} $$ where $n_4$ is the number density ($n$) of
single stars in units of $10^4~ \rm{pc^{-3} }$ and $v_{10}$ is
the velocity dispersion ($v$) in units of 10 km/sec in GCs;
$P_{orb}$ is the orbital period in days giving $t_{fly}$ in years.  The value of $v_{10}/n_{4}$ (in GC center) varies from 0.0024 - 2.167 for different GCs \cite{web85}. We grouped the GCs into six groups according to the values of $v_{10}/n_{4}$ and calculated $t_{fly}$ with the mean values of $v_{10}/n_{4}$ for
each group. In Fig. \ref{fig:pulsars_all_group}, we plot the isochrones of fly-by encounters ($t_{fly}$) in the $e - P_{orb}$ plane for all groups.  If $t_{fly} > 10^{10}$ years for a particular binary, then it would not have interacted and it would
preserve its original eccentricity. If $t_{fly} < 10^{10}$ years
for a particular binary, then it could be eccentric due to fly-by
interactions which is the case for most of the eccentric GC binaries (Fig. \ref{fig:pulsars_all_group}). Therefore, many GC pulsars' orbital eccentricities can be explained by fly-bys. However the local stellar densities at the pulsar positions (especially with positional offset from cluster cores) may result higher values of $v_{10}/n_{4}$ from the central values used here. Thus these pulsars might be in $t_{fly} > 10^{10}$ years region. Moreover, for pulsars with $e > 0.1$ (a majority of the group I pulsars),  very close fly-bys are necessary if the initial binaries were circular (Camilo \& Rasio 2005). In these cases, exchange and merger events are relevant. But the three pulsars with $0.01< e <0.1$ and $60 < P_{orb} < 256 \; \rm d$ are possibly white dwarf cores of red giant companions that overflowed Roche lobe \cite{web83}. Such binaries would normally have the``relic" eccentricities $\sim 10^{-4}$ (Phinney 1992). The above binary pulsars with their presently mildly high eccentricities, have undergone fly-by encounters with single stars, rather than exchange reactions, which would produce very high eccentricities $e > 0.1$. 

\begin{figure}[h!]
\plottwo{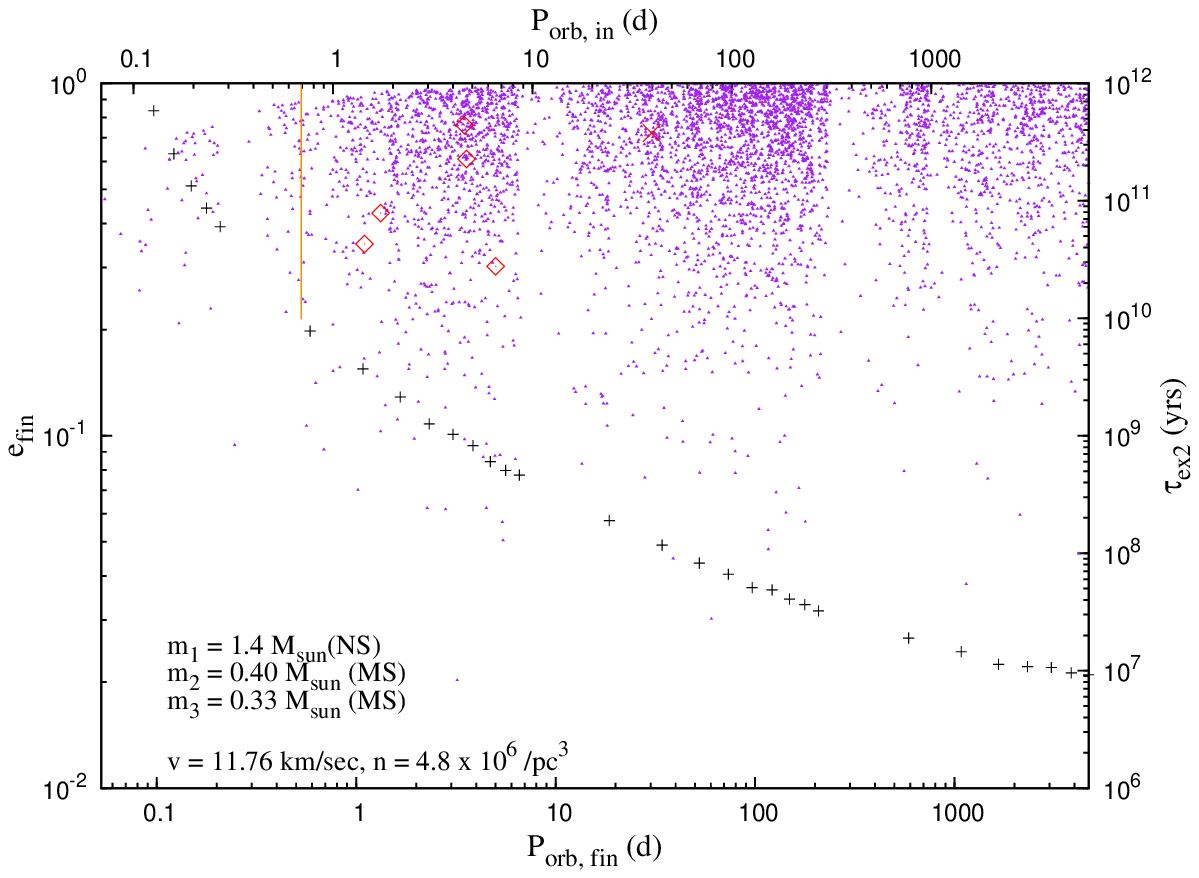}{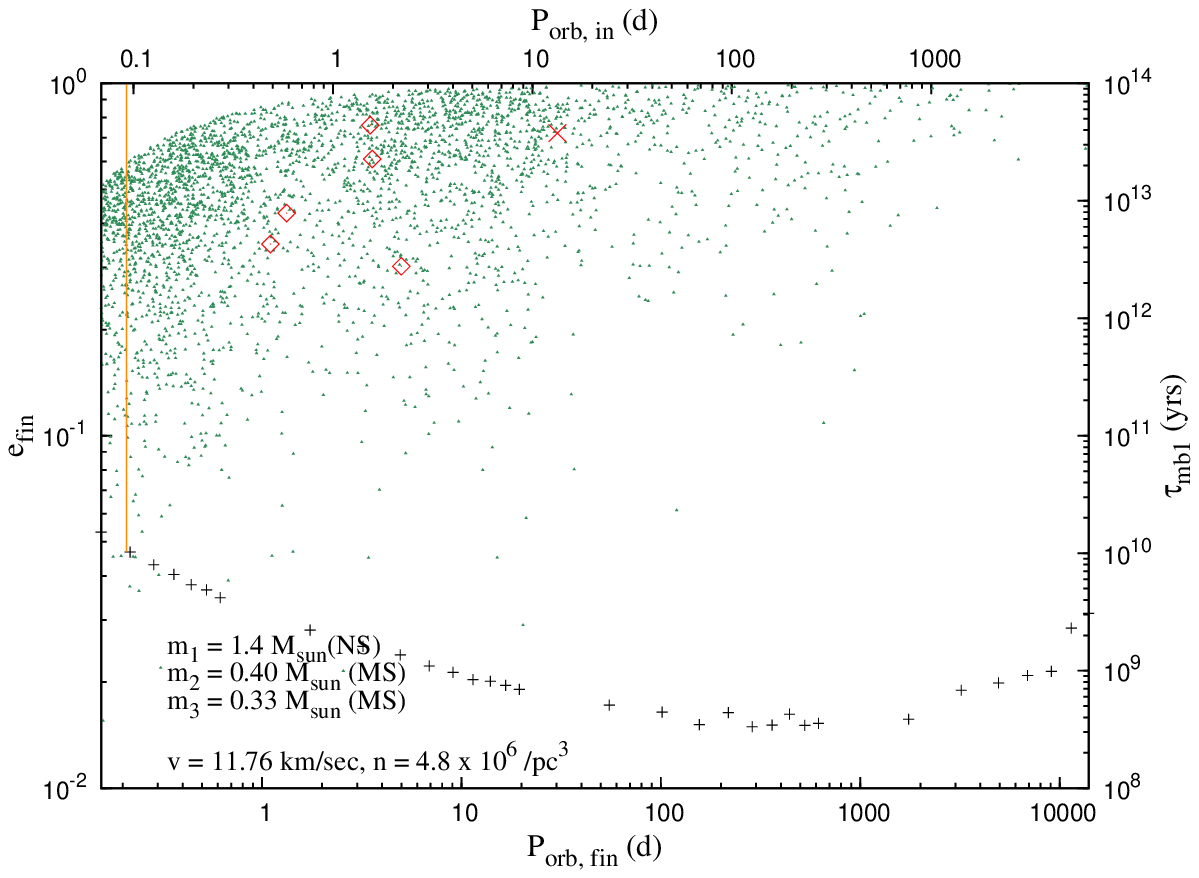}
\plottwo{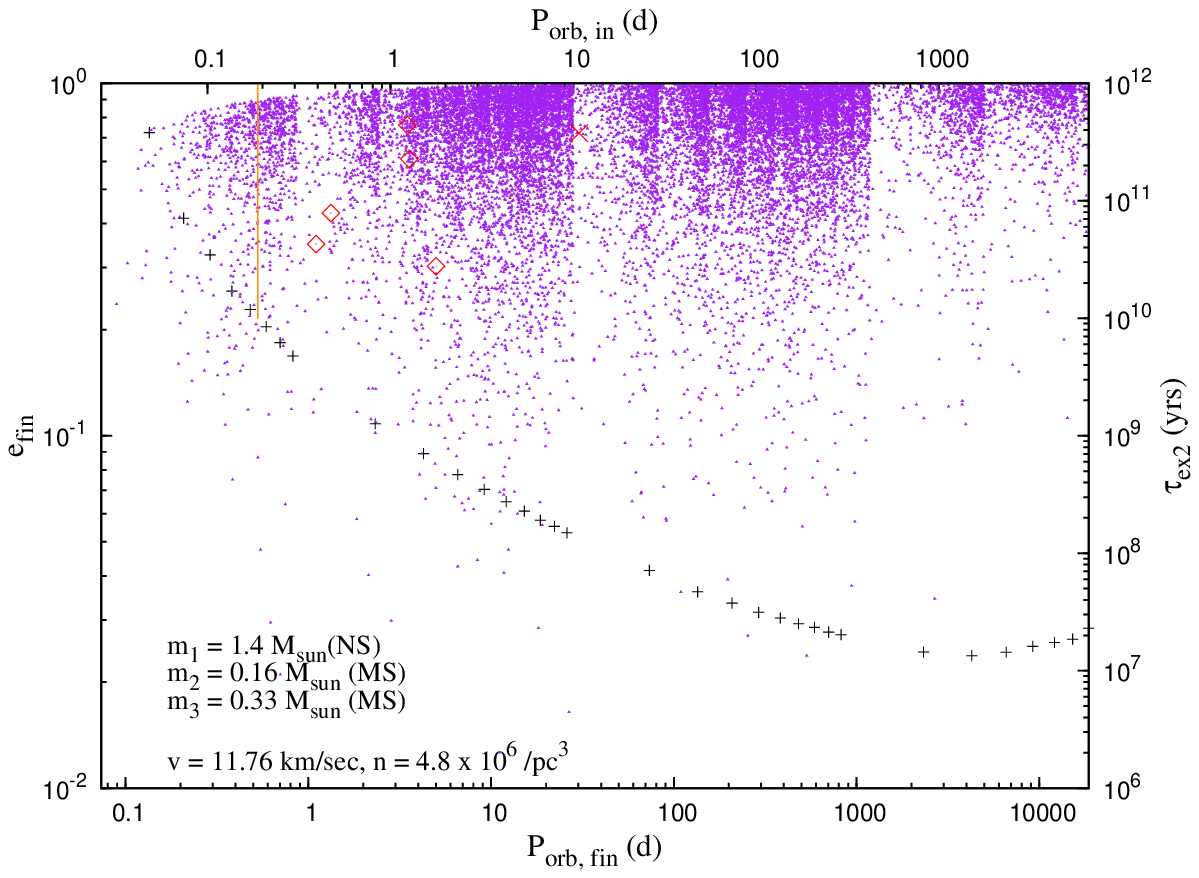}{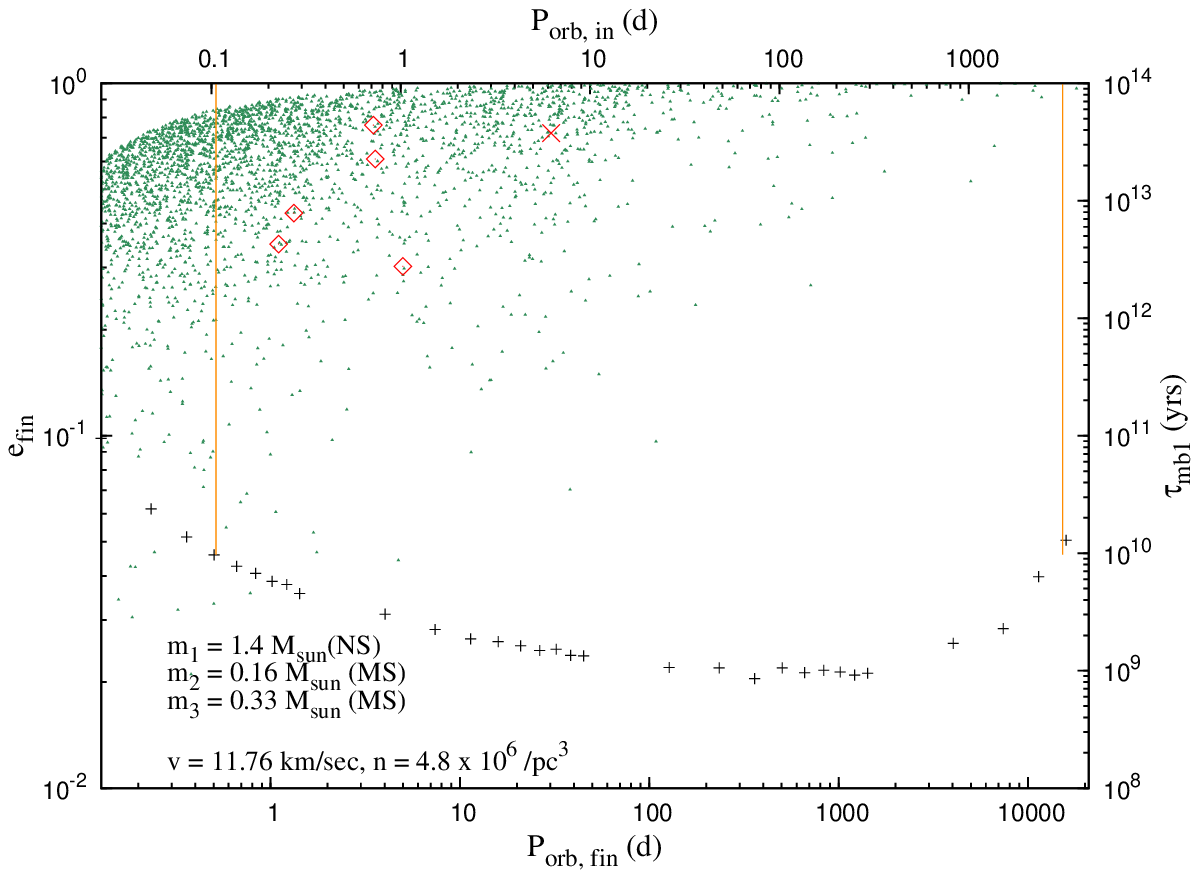}
 \caption{Time scales (denoted by black `+') and final
eccentricity distributions (colored scatter plots) with initial
and final orbital periods for exchange (purple points on the left panel) and merger (green points on the right panel) interactions with different stellar parameters. 
Vertical orange lines give boundaries of orbital periods where
interaction time scales $ < 10^{10}$ yrs. 
\label{fig:terzan_exch_merg} }
\end{figure}

We used STARLAB\footnote{www.ids.ias.edu/$\sim$starlab/}
codes to perform numerical simulations of exchange and merger with different stellar parameters (GC parameters same as Terzan 5, since it has the lowest value of the parameter $v_{10}/n_4$ where all classes of encounters can take
place).  In Fig. \ref{fig:terzan_exch_merg}, the left y axis gives the final eccentricities
while the right y axis gives the time scales of interactions. Moreover, we plot
$P_{orb, in}$ of the initial binary along the top x-axis and $P_{orb, fin}$  along the
bottom x-axis. $P_{orb, fin}$ is obtained from $P_{orb, in}$
putting $\Delta=0$ in the relation $a_{fin}=\left[a_{in}m_a m_b/m_1 m_2 \left(1-\Delta \right) \right]$ where $m_1$ and $m_2$ are masses of the members of the initial binary, $m_3$ is the mass of the incoming star, $m_a$ and $m_b$ are masses of the members of the final binary, $\Delta$ is the fractional change of binary binding energy. For exchange, $m_a=m_1$, $m_b=m_3$; for merger, $m_a=m_1$, $m_b=m_2+m_3$. Six high eccentricity ($e > 0.1$) binaries in Terzan 5 are also shown (red in colors, symbols same as used in Fig \ref{fig:pulsars_all_group}) here. It is clear from the scatter plots (Fig. \ref{fig:terzan_exch_merg}) that the final binaries will most probably have $e>0.1$ if they undergo either exchange or merger events. If a pulsar lies in a region where $t_{int}>10^{10}$ years in a particular plot, then that pulsar can not experience that interaction. Moreover, it the companion masses are much different from expected values ($m_3$ for exchange and $m_2+m_3$ for merger) then that interaction should also be excluded. Thus we conclude that PSR I (in Ter5) can be either result of exchange if $m_2$ was either $0.16~M_{\odot}$ or $0.40~M_{\odot}$ or merger if $m_2$ was $0.16~M_{\odot}$, PSR J and X can be results of exchange if $m_2$ was either $0.16~M_{\odot}$ or $0.40~M_{\odot}$, PSR Q, U and Z can be results of merger only if $m_2$ was $0.40~M_{\odot}$ where in all cases we consider $m_3=0.33~M_{\odot}$ (average mass of single stars in GCs).

Thus we conclude that mildly eccentric binaries could have been generated by fly-by encounters with low eccentricity pulsars below the line of ``timing sensitivity limit" (group III pulsars). Some of the shorter $P_{orb}$ binaries
would again be circularised by gravitational radiation (see $t_{gr}$ contours in Fig \ref{fig:pulsars_all_group} calculated using the formalism of Peters \& Mathews 1963).
Group III binaries lie in the region $t_{fly}< 10^{10}$ years but they have not get eccentricities. Gravitational wave radiation can explain the circularity of tighter binaries among them, but others are either very young or lie far away from the cluster centre where $v_{10}/n_{4}$ has higher values. Extremely high eccentric binaries are probably results of either exchange or merger interactions. See Bagchi \& Ray (2009) for details.

\end{document}